# Observation of robust edge superconductivity in Fe(Se,Te) under strong magnetic perturbation

Da Jiang[1#*], Yinping Pan[2#], Shiyuan Wang[2#], Yishi Lin[2], Connor M. Holland[3], John R. Kirtley[3], Xianhui Chen[4], Jun Zhao[2], Lei Chen[1], Shaoyu Yin[5, 2*], Yihua Wang[2, 6*]

1. Shanghai Institute of Microsystem and Information Technology, Shanghai, 200050, China.
2. State Key Laboratory of Surface Physics and Department of Physics, Fudan University, Shanghai 200433, China.
3. Department of Physics, Stanford University, Stanford, CA 94305, USA.
4. Department of Physics, University of Science and Technology of China, Hefei 230026, China.
5. Institute for Theoretical Physics and Cosmology, Zhejiang University of Technology, Hangzhou 310023, China.
6. Shanghai Research Center for Quantum Sciences, Shanghai 201315, China.

# These authors contributed equally to this work.

* To whom correspondence and requests for materials should be addressed. Email: jiangda@mail.sim.ac.cn; syyin@zjut.edu.cn; wangyhv@fudan.edu.cn

**The iron-chalcogenide high temperature superconductor Fe(Se,Te) (FST) has been reported to exhibit complex magnetic ordering and nontrivial band topology which may lead to novel superconducting phenomena. However, the recent studies have so far been largely concentrated on its band and spin structures while its mesoscopic electronic and magnetic response, crucial for future device applications, has not been explored experimentally. Here, we used scanning superconducting quantum interference device microscopy for its sensitivity to both local diamagnetic susceptibility and current distribution in order to image the superfluid density and supercurrent in FST. We found that in FST with 10% interstitial Fe, whose magnetic structure was heavily disrupted, bulk superconductivity was significantly suppressed whereas edge still preserved strong superconducting diamagnetism. The edge dominantly carried supercurrent despite of a very long magnetic penetration**

**depth. The temperature dependence of the superfluid density and supercurrent distribution were distinctively different between the edge and the bulk. Our Heisenberg modeling showed that magnetic dopants stabilize anti-ferromagnetic spin correlation along the edge, which may contribute towards its robust superconductivity. Our observations hold implication for FST as potential platforms for topological quantum computation and superconducting spintronics.**

Iron-chalcogenide Fe(Se,Te) (FST) is an important family of iron-based high transition temperature ($T_c$) superconductors (FeSC) [1] [2] [3] [4] [5] [6] [7] [8] [9] [10]. Despite of its simple crystal structure, monolayers of FST grown on SrTiO$_3$ have shown the highest $T_c$ among all FeSC's [11] [12] [13] [14]. Stronger spin-orbit coupling with increasing Te concentration in FST could lead to topologically non-trivial band-inversion in both the bulk FST crystals [15] [16] [17] [18] and monolayers grown on SrTiO$_3$ [11] [12] [13] [19] [20]. Scanning tunneling spectra have even shown signs of zero-bias peaks at vortex cores [21] [22] [23] and interstitial Fe sites [24], which were considered as a signature of Majorana fermions. While the electronic structures are similar across FeSC [2], the magnetic structure of FST, which is believed to intimately influence the superconducting pairing [25] [26] [27], depends on the chemical composition of Se/Te and more sensitively on interstitial Fe [6] [7] [8] [9] [10].

All these investigations of the local band and spin structure by spectroscopy and scattering techniques are highly suggestive of emergent electron-magnetic phenomena on a mesoscopic scale [27]. Just like the boundary states of topological insulators and quantum anomalous Hall insulators being robust against non-magnetic impurity scatterings, non-trivial topology of a bulk superconductor could also manifest itself through protected boundary supercurrent [28]. Nevertheless, how the boundary superconductivity is affected by the ubiquitous magnetism in FST is by no means apparent and yet it may provide a critical step in our study of the interaction between superconductivity and magnetism of FST. Therefore, directly distinguishing the edge or surface superconductivity from the bulk and determining its superfluid density and supercurrent density under strong magnetic perturbation may have compounded implications for both fundamental understanding of unconventional superconductivity and any future applications in superconducting spintronics [29] [30] and topological quantum computation on FST platforms. However, mesoscopic electro-magnetic measurements essential towards these understandings and applications are still lacking as they require a sensitive magnetic imaging technique capable of distinguishing between bulk and edge superconductivity [31].

Here, we used scanning superconducting quantum interference device (sSQUID) microscopy in magnetometry, current and susceptibility modes [32] to image the supercurrent distribution and local superfluid density in superconducting FST with excess Fe. Interestingly, we found that highly diamagnetic regions existed only along the edge of the sample while the bulk showed weak diamagnetism with much reduced $T_c$ due to the suppression effect from interstitial Fe. The edge also dominantly carried the supercurrent biased into the sample, further suggesting the existence of independent superconducting channel with high critical current density along the edge. Our result was a clear indication that the edge superconductivity in FST was robust in the sense that it was not suppressed by interstitial Fe.

The crystal structure of FST is composed of the primitive tetrahedral Fe-anion cage (Fig. 1a) on which all FeSC's are based [1] [2] [3]. The Fe atoms inside the cages (labelled as Fe1 in Fig. 1a) form a square-lattice plane which constitutes the basis of studying the electronic and magnetic structure of all FeSC's. The Fe1 plane is sandwiched between two anion (Se/Te) planes, and such trilayer structure repeats in the $c$ direction with very weak coupling between layers. The excess Fe atoms occupy interstitial sites in the Se/Te plane (labelled as Fe2 in Fig. 1a). The superconductivity and magnetic structure of FST evolve with its chemical composition [9]. With increasing Se doping of the parent compound FeTe, a long-range anti-ferromagnetic order is suppressed and superconductivity emerges [8]. The interstitial Fe spins have been found through neutron scattering measurements to perturb the magnetic structure of the Fe1 plane [7] [8] and significantly reduce the superconducting volume fraction in the superconducting samples [3]. Despite their suppressed bulk superconductivity, $T_c$ of such "non-bulk" superconducting FST samples with excess Fe was found through transport measurement to be very similar to their bulk superconducting counterparts [4] [5] across the phase diagram [1].

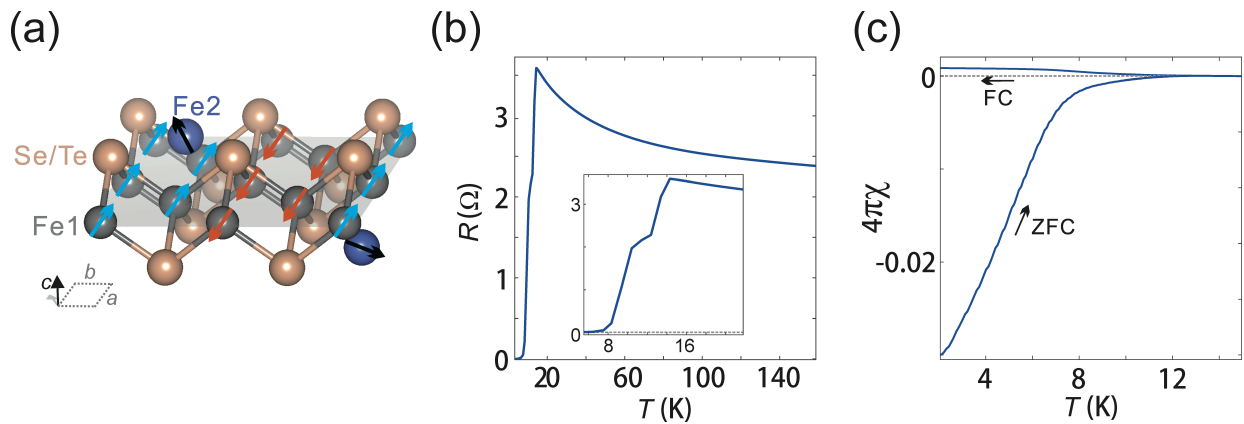

**Fig. 1** (Color online) Suppressed bulk superconductivity in a Fe-chalcogenide superconductor with excess Fe. (a) Crystal structure of Fe(Se,Te) (FST). Fe1 atoms (gray), bonding with the nearest Se/Te (brown) in a tetrahedron, form a planar square lattice, which is sandwiched between the Se/Te planes. Such a structure repeats in the $c$ direction. The excess Fe (Fe2, purple) occupies the interstitial sites in the Se/Te plane. Arrows on Fe1 (blue, aligned in-plane) indicate the diagonal double stripe antiferromagnetic spin structure of its parent compound FeTe. Fe2 spins (black arrows) disrupt such an order. (b) Typical resistance as a function of temperature ($T$). The increasing normal state resistance as a logarithmic function of decreasing $T$ is a sign of carrier localization. (b) inset: a zoom-in of the low $T$ section showing the transition starting around 14 K. (c) Magnetization vs. $T$ curves for no-bulk $Fe_{1+y}Se_{0.5}Te_{0.5}$. "FC" and "ZFC" are the field cooled and zero-field cooled curves, respectively, where the field was 10 Oe and normal to the Fe1 plane.

In this study, we concentrate on the sample compositions of $Fe_{1.11}Se_{0.40}Te_{0.60}$ (FST04) and $Fe_{1.07}Se_{0.47}Te_{0.53}$ (FST05) which are within the "non-bulk" superconducting regime [3] [4] [8]. The Se/Te ~ 1:1 is relevant to some recent spectroscopic studies where nontrivial band-structure was found [13] [18] [21] [24]. The slight variance in composition did not affect their superconducting behavior very much, as will be seen below. Their normal state resistance rose as the temperature ($T$) decreased (Fig. 1b), suggesting localization of the carriers [8]. The resistance started to decrease sharply at 14 K (Fig. 1b inset) in agreement with the $T_c$ from other transport measurements [4] [5]. However, diamagnetic susceptibility did not show a sudden jump upon entering the superconducting state as would be expected in a bulk superconductor, indicating depression of bulk superconductivity [3] [4] [8] [33]. Superconducting volume fraction increased faster below 8 K as revealed by the diamagnetic susceptibility (Fig. 1c) although it only reached 3% at the base $T$. This was consistent with previous results [3] and indicated a very weak superconductor far from full Meissner state. Our extensive spatially-resolved characterizations of these sample, including state-of-the-art scanning transmission electron microscopy and confocal Raman spectroscopy, suggested that the chemical composition and lattice structure were uniform in the mesoscopic scale relevant for this study (see the Supplementary materials).

To visualize the spatial distribution of superfluid density in FST, we employed scanning magnetic imaging with a nano-scale SQUID susceptometer [34] [35] [36]. The pickup loop was integrated into a two-junction SQUID that converted the flux through the loop ($\Phi$) into a voltage signal for direct magnetometry. A current ($I_F$) in the field coil provided a local magnetic field for susceptometry measurement ($d\Phi/dI_F$). Current magnetometry ($\Phi'_I/I_{AC}$) measured the in-phase component of the flux ($\Phi'_I$) through the pickup loop in response to an alternating current (AC) source-drain bias current ($I_{AC}$). The spatial resolution throughout this work was around 4 μm (see the Supplementary materials).

Susceptometry and magnetometry performed on an exfoliated FST04 sample enabled a direct visualization of the spatial pattern of superconducting diamagnetism (Fig. 2). To prevent any oxidation effect, we capped the 700-nm-thick sample with a layer of graphene right after exfoliation, which did not affect the scanning SQUID measurement. The bulk of the sample showed very weak diamagnetism for a sample of such thickness, while that along the edge of the sample was relatively stronger (Fig. 2b). The weak bulk diamagnetism in FST with excess Fe could be attributed to Fe2 spins disrupting the phase coherence of superconducting order parameter in the bulk. In comparison with the left and the bottom edges, the overall weakness of the top edge diamagnetism (though still stronger than the bulk) was a susceptometry artefact (see the Supplementary materials I.3) dependent upon the orientation of the leads of the pickup loop and field coil (Fig. 2b and Fig. S4b online, smaller and larger coils respectively). Nonetheless, the local inhomogeneity of intensity along a certain edge was intrinsic. As control experiments, we found that the bulk superconducting FeSe and FeSe$_{0.4}$Te$_{0.6}$ and the non-bulk superconducting Fe$_{1.12}$Se$_{0.2}$Te$_{0.8}$ did not show distinct edge feature in susceptometry (Fig. S3 online). The dependence of enhanced edge diamagnetism on the chemical composition of FST rules out instrumental artefacts and extrinsic effects such as oxidation and topographic inhomogeneities along the edge (more discussions in the Supplementary materials II).

We did not observe any isolated vortices in the bulk of the sample from the magnetometry image (Fig. 2c). This was atypical because there were residual magnetic fields in the measurement setup on the order of mG, which would lead to easily observable trapped vortices for sSQUID to detect in a conventional superconductor. Instead, the magnetic contrast was mainly along the edges and step edges of the sample where the diamagnetic contrast was dominant (Fig. 2b and c). The two dipole-like features in magnetometry (Fig. 2c), possibly due to some magnetic dust, were absent in susceptometry (Fig. 2b), suggesting that our magnetometry and susceptometry were not interdependent. The coincidence of the strong magnetic and diamagnetic contrast along the edges (Fig. 2b and c) was therefore nontrivial and could be relevant to the origin of the edge superconductivity.

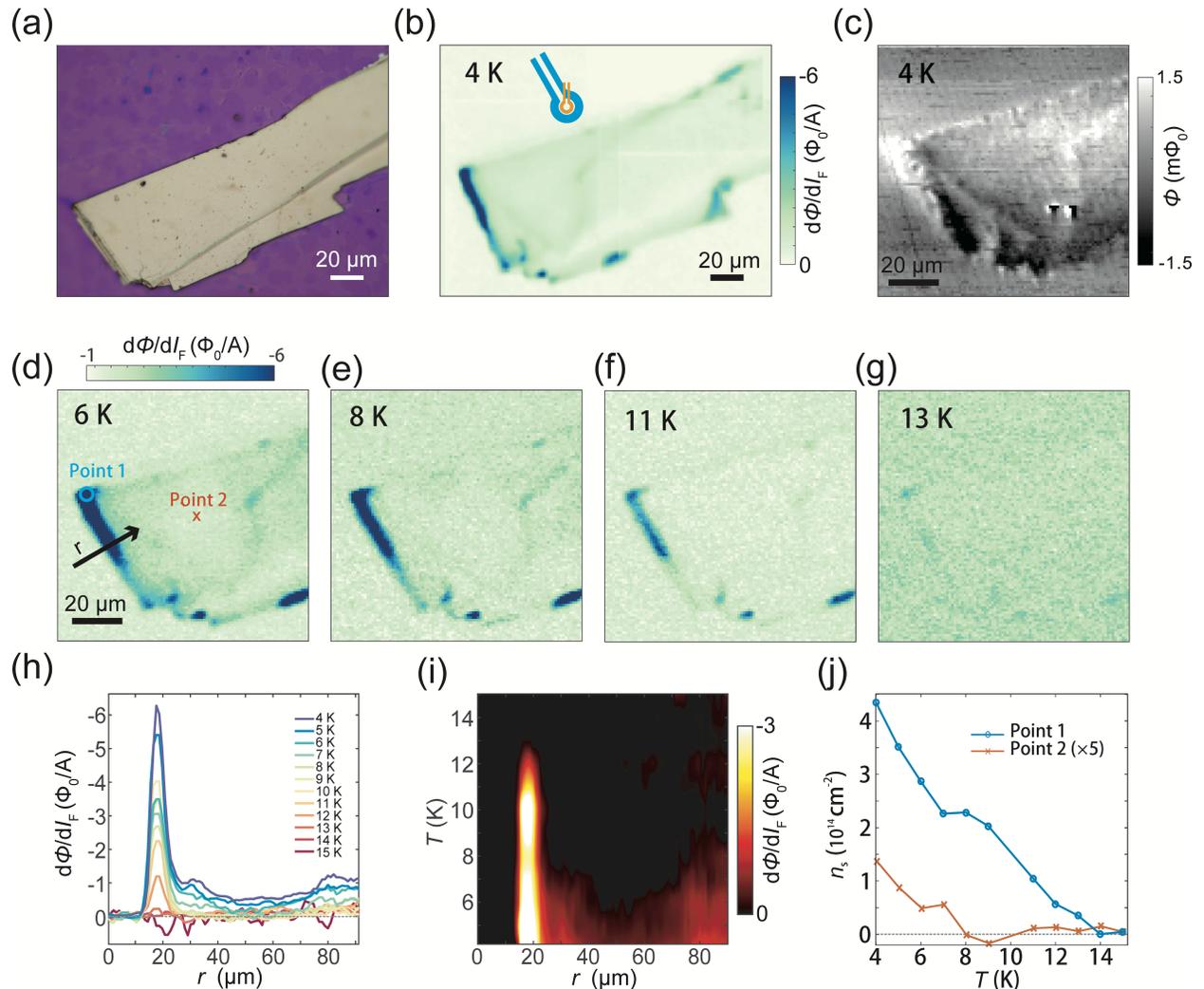

**Fig. 2** (Color online) Distinctive edge features in susceptometry of sample FST04 ($Fe_{1.11}Se_{0.40}Te_{0.60}$) with graphene capping. (a) Optical image of the FST04 sample. The width of the sample is about 100 μm. (b) and (c) are the susceptometry and magnetometry images of the sample, respectively. The relative orientations of the SQUID pickup loop (smaller orange one) and field coil (larger blue one) with respect to the sample are illustrated in (b). (d–g) Susceptometry images of the sample at various $T$. (h) line cuts of the susceptometry images at various $T$ along the vector direction ($r$) as labeled by the arrow in (d). Zero is at the starting position of the arrow and the cuts extend across the sample. (i) Interpolated image from the line-cuts in (h). (j) Superfluid densities as a function of $T$ extracted from point 1 and 2 in (d). Note that the data for point 2 is enlarged by a factor of 5.

Both the edge and bulk diamagnetism became weaker as the $T$ increased (Fig. 2d–g), which was expected. However, edge and bulk showed different $T$ dependence. We took line cuts perpendicular to the left edge of the sample (arrow direction in Fig. 2d) at various $T$ to better visualize their distinction (Fig. 2h and i). The diamagnetic response of the bulk was mostly flat at 4 K (Fig. 2h, purple) and it reduced quickly as $T$ increased (Figs. 2h and i). Above 8 K (Fig. 2i) the diamagnetic response of the bulk was indistinguishable from that above the $T_c$ (Fig. 2i). In stark contrast, the diamagnetism of the left edge at 4 K showed up as a strong peak (Fig. 2h and i). This peak reduced in magnitude as $T$ increased and persisted till 13 K, above which it was indistinguishable from the background (Fig. 2i). The full width half maximum of the peaks did not vary with $T$ and were about 5 μm up to 12 K (Fig. 2h and i), which was very close to our spatial resolution.

According to the Ginzburg-Landau theory of an inhomogeneous superconductor [33], there are two different characteristic length scales which determine the variation of pairing wave-function and magnetic field, respectively. The pairing wave-function (the square of which is superfluid density) is determined by the coherence length and the magnetic field by London penetration depth. For a weak planar superconductor, diamagnetic AC susceptibility is proportional to the superfluid density and therefore the features in susceptometry intrinsically vary with the coherence length. Since the coherence length of FST is on nanometer scale at low $T$, the width of the edge diamagnetism we measured (Fig. 2h) was mainly limited by sSQUID's resolution (until within 1 ppm of $T_c$, which we cannot detect due to much reduced signal-to-noise ratio). On the other hand, the spatial variation of the magnetic features in a thin superconductor, whose thickness is smaller than the London penetration depth, is determined by the Pearl length. Magnetic features such as those from vortices and Meissner current in a conventional type II thin superconductor, which we visualize through magnetometry, are typically larger than our resolution and thus enlarge with increasing $T$.

Having discussed the spatial variation of the edge diamagnetic susceptibility, we now turn to its magnitude. As mentioned earlier, it is proportional to the inverse of the Pearl length [33], which allowed us to extract the superfluid density as a function of $T$. Using the Pearl length obtained from the touch-down curve, we extracted the superfluid density $n_s$ (see the Supplementary materials) from susceptometry at different locations corresponding to the edge and the bulk (Fig. 2j). Based on the obtained edge and bulk $n_s$, we gave a rough estimate of the width of edge channel to be 5 nm (see the Supplementary materials). The edge $n_s$ dropped in a linear fashion as $T$ was increased to 14 K (Fig. 2j, blue circles). In agreement with the line-cuts (Fig. 2h), $n_s$ of the bulk was much smaller and disappeared below the sensitivity limit of our detection at around 8 K (Fig. 2j, orange crosses). This temperature was also consistent with the change of slope in the superconducting volume fraction vs. $T$ as measured in bulk magnetometry (Fig. 1c), indicating that 8 K could be taken as the bulk $T_c$. The fact that the edge had a distinctly higher superfluid density and $T_c$ than those of the bulk indicated that their superconductivity was affected differently under the strong magnetic perturbation. Given the edge $T_c$ being the same as that of bulk superconducting FST, we could draw a conclusion that phase coherence on the edge was not as significantly suppressed by Fe2 as in the bulk [8] [10].

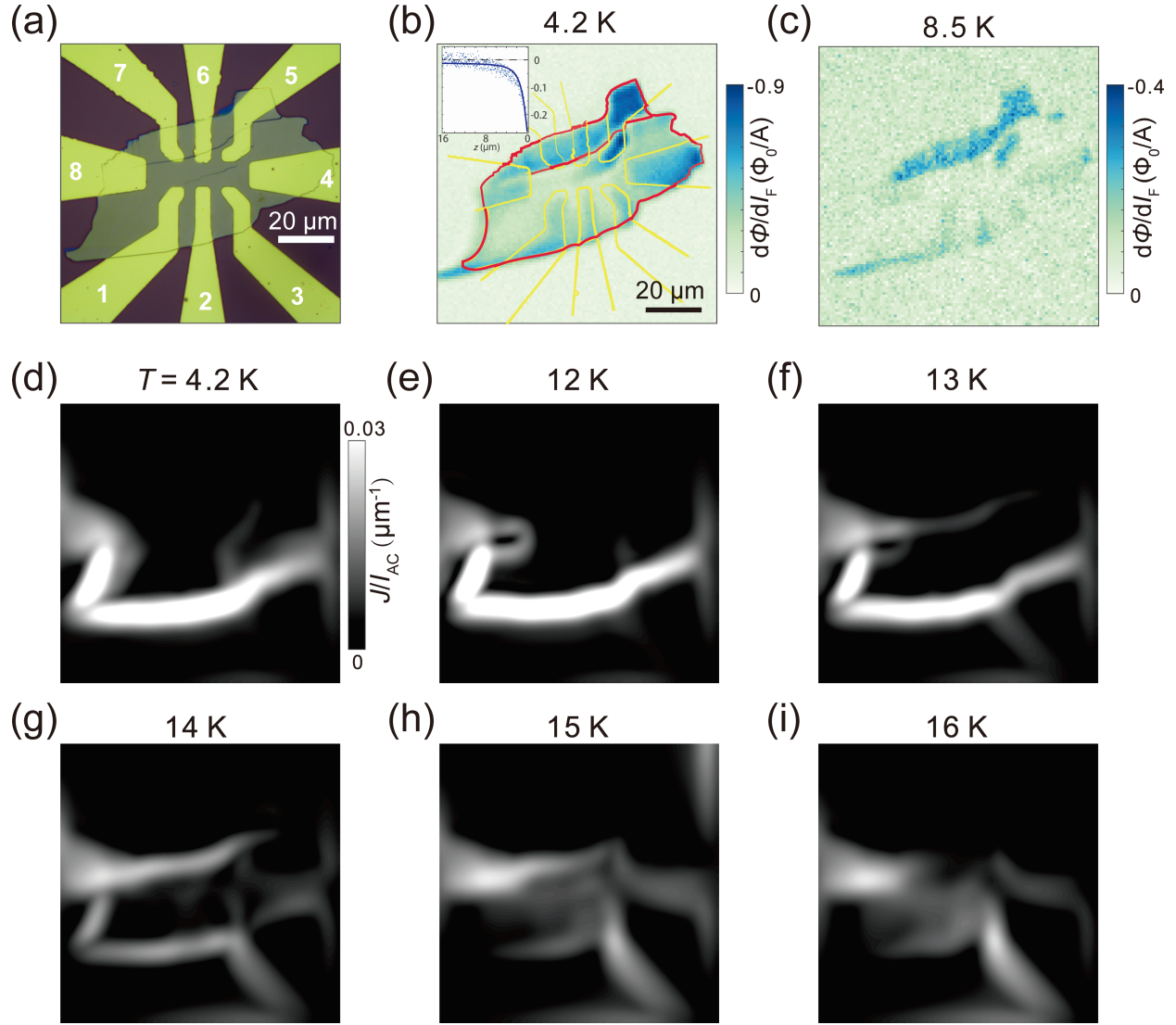

**Fig. 3** (Color online) **Diamagnetism and edge current flow in a FST05 (Fe$_{1.07}$Se$_{0.47}$Te$_{0.53}$) sample.** (a) Optical image of the sample flake with Ti/Au electrodes (patterns with numbers). The applied current of 0.1 mA is between electrodes 8 and 4. (b, c) Susceptibility images taken at 4.2 and 8.5 K, respectively. The edges of the sample and the bulk areas under the electrodes show stronger diamagnetic signals than other areas of the bulk. The outline of the sample as well as the crack lines from the optical image are overlaid with red lines, those of the electrodes with yellow lines. (b) inset: a touch-down curve of susceptibility on the left edge at 4.2 K. (d–i) Current flow images from the device shown in (a) at the labeled $T$, showing the evolution from edge (supercurrent) conduction to bulk (normal) conduction.

The susceptometry and current magnetometry images (Fig. 3) on an FST05 sample patterned with electrodes showed that the edge was mainly responsible for the $T_c$ we observed from transport (Fig.

1b). Similar to the FST04 sample (Fig. 2), this much thinner sample about 60 nm thick (Fig. 3a) also displayed very weak diamagnetism in the bulk region and relatively stronger diamagnetism along its edge (Fig. 3b and c). That was to the exception of the regions covered with the Ti(10nm)/Au(100nm) electrodes (Fig. 3a), which showed significant enhancement of diamagnetic response (Fig. 3b). Due to the finite resolution of our sSQUID, gap areas between the electrodes 5 μm apart were also smeared by signal from both the edge and the electrodes. Such extrinsic enhancement at the electrode area was introduced during the nano-fabrication process and it was absent on the samples fabricated with a different technique (see the Supplementary materials). The touch-down curve (Fig. 3b inset) showed local diamagnetic signal of this sample on the edge at 4.2 K corresponding to a Pearl length of $\Lambda \sim 3$ mm, whereas that of the bulk was $\sim 12$ mm [37]. Such long Pearl length comparing with the size of the sample explained the absence of isolated vortices in the magnetometry image (Fig. 2c). The supercurrent flowing along the edge had a much smaller characteristic width than the Pearl lengths and therefore cannot be due to the Meissner effect [33].

Contrary to a uniform current distribution which an extremely long Pearl length would have resulted in, current magnetometry images (Fig. 3d–h) showed that current was localized on the lower edge of the sample at $T = 4.2$ K (Fig. 3d). A crack line was visible separating the upper-right corner of the sample (Fig. 3a), which might have caused finite contact resistance that prevented the current from flowing to lead 4 (Fig. 3a and b) via the top edge (Fig. 3d). As the bulk superfluid density was not completely suppressed at this temperature (Fig. 3b, also conf. Fig. 2j), such a current distribution indicated that the applied current of 0.1 mA was well below $I_c$ of the edge but higher than that of the bulk at 4.2 K. The current distribution did not change much as $T$ was increased to 12 K (Fig. 3e), at which temperature the bulk superconductivity was already suppressed as shown in the susceptometry images at 8.5 K (Figs. 3c and 2g). The current distribution changed more drastically as $T$ approached $T_c$ (Fig. 3f–i) with a branch of current flowing through the crack line on the upper part of the sample at 13 K (Fig. 3f and a). This was consistent with the appearance of finite resistance on the edge right below $T_c$ (Fig. 1b inset). The edge current density significantly reduced at $T = 14$ K (Fig. 3g) and above 15 K the current flow became diffusive (Figs. 3h and i). This coincided with the temperature at which the entire sample became normal (Fig. 1b inset).

In our device for current magnetometry, the electrodes cover both the edge and the bulk of the sample connected in parallel. Their decoupled conducting behavior again revealed that the characteristics of superconductivity were different between the edge and the bulk. From our transport measurement, a finite resistance was still present below 8 K (Fig. 1b inset) when the bulk was weakly diamagnetic and the edge showed stronger diamagnetism (Fig. 2j). There was also a step around 9 K in the resistance-temperature curve (Fig. 1b inset), which was reminiscent of transport behaviors in superconducting wires [33]. These effects suggested strong superconducting fluctuation on the edge and hinted the low dimensional nature of the edge superconductivity [38] [39] [40] [41]. Such fluctuation effects prevented the edge from shorting out the entire sample below 15 K and played an important role in the charge transport (Fig. 1b) and current distribution (Fig. 3f and g). The relatively strong coupling between Fe-layers likely stabilized the edge superconductivity so that diamagnetism persisted to 15 K despite of the fluctuations.

As the sSQUID measurements sum up magnetic contributions over the sample thickness, the diamagnetic signal of the top and bottom surfaces is part of the bulk response. The absence of superconductivity in the bulk above 8 K suggested that the two surfaces were not more superconducting than the interior. In contrast, the edge showed much larger superfluid density and critical current with higher $T_c$. We thus infer that the robustness of the edge superconductivity stemmed from being on the boundary of 2D FST layers instead of being on the (side) surface of a 3D crystal. Considering the recent observation of nontrivial band inversion and superconductivity in monolayer FST [13] [14] [42], our result suggests monolayer FST with excess Fe may be a candidate for chiral 2D topological superconductor [16] whose boundary inevitably hosts protected edge states, i.e., Majorana modes. Although the topological nature of bulk FST with excess Fe is not clear at this point, it is possible that their edge might exhibit high-order topological structure [43] [44] with broken time-reversal symmetry [45].

Besides breaking time-reversal symmetry of the topological band-structure, interstitial Fe2 spins also have a non-trivial effect on the Fe1 spin-structure in the square lattice (Fig. 1a). Recent work showed that Fe2 spins in the bulk favored collinear alignment with their nearest-neighbor spins [6] and they did not affect the amplitude of local superconducting gap [24] [46]. There have been both theoretical proposal [2] [25] [26] [27] and experimental evidence which associated antiferromagnetic spin correlation to superconducting pairing in iron-based superconductors [8] [9] [10]. From these studies, we can infer that randomly polarized Fe2 spins, by reorienting their nearest-neighbors into a ferromagnetic configuration, cost Fe1 spins short range antiferromagnetic correlation and possibly long-range superconducting phase coherence [39]. The spatial dependence of the spin-correlation found in those studies motivated us to investigate whether Fe2 moment could interact with the spins located on the edge and the bulk differently.

To this end, we have performed numerical simulation based on the Heisenberg model (Fig. 4 and the Supplementary materials) that have been adopted for iron-chalcogenides previously [6] [27] [47]. We took into account nearest- and next-nearest neighbors spin exchange couplings under both open (Fig. 4a) and periodic (Fig. 4b) boundary conditions. The average antiferromagnetic correlation between Fe1 spins ($C_{\langle 11 \rangle}$) varied as a function of anti-ferromagnetic coupling coefficient between nearest Fe1 moments ($J_{11}$); while the ferromagnetic coupling between the nearest Fe2 and Fe1 ($J_{21}$) was chosen to be the energy scale and was set to $-1$. The next-nearest-neighbor coupling between Fe1 spins ($J_{12}$), though absent in the edge chains due to geometry (Fig. 4), was included in the bulk cases (see the Supplementary materials for the full model).

As a function of increasing Fe2 perturbation (decreasing $J_{11}$), the simulation showed that $C_{\langle 11 \rangle}$ was reduced by the Fe2 spins both on the edge and in the bulk (Fig. 4a and b) as expected. $C_{\langle 11 \rangle}$ of the edge was systematically lower than that of the bulk, indicating a stronger antiferromagnetic correlation on the edge. The most striking result was that while bulk antiferromagnetic $C_{\langle 11 \rangle}$ was quickly suppressed with increasing strength of the Fe2 perturbation in the regime of $0.7 < J_{11} < 1$ (the most relevant parameter range for FST), antiferromagnetic $C_{\langle 11 \rangle}$ of the edge varied very slowly (Fig. 4a and b). The results are qualitatively consistent regardless of the boundary condition and simulating parameters such as edge length. All the above were strong indications that edge antiferromagnetic spin correlation was protected against the Fe2 perturbation.

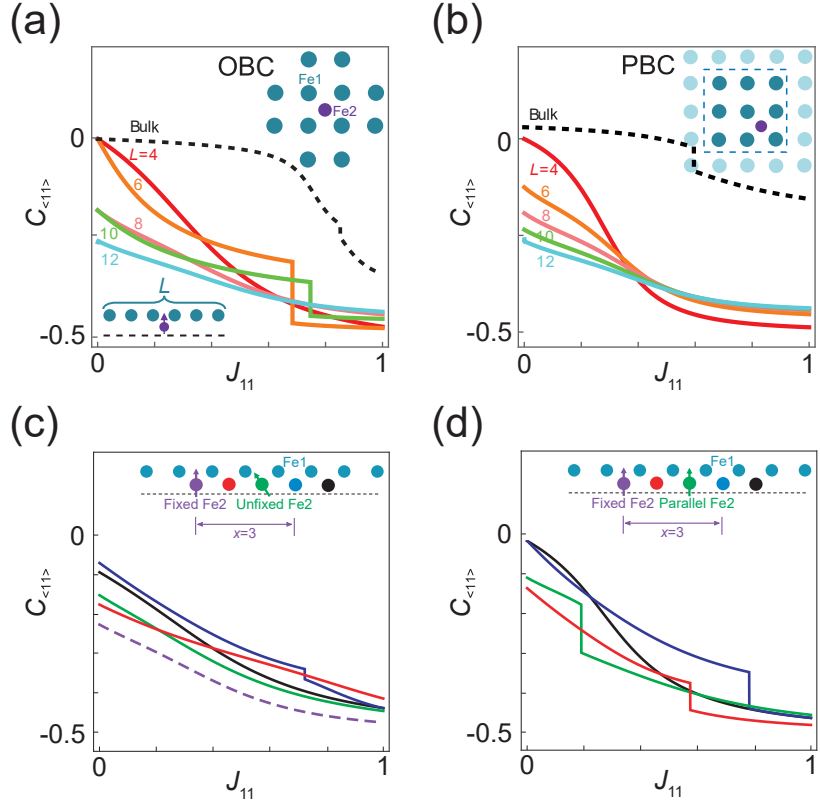

**Fig. 4** (Color online) Heisenberg modeling showing stronger anti-ferromagnetic (AFM) spin correlation along the edge than in the bulk under magnetic perturbation. (a) and (b) are spin correlation between nearest-Fe1 moments ($C_{\langle 11 \rangle}$) in presence of a single Fe2 moment for open boundary condition (OBC) and periodic boundary condition (PBC), respectively, as a function of $J_{11}$, the exchange coupling coefficient between nearest-Fe1 moments. For other parameters in our full model: between Fe2 and its nearest-Fe1 moments, the coupling coefficient $J_{21} = -1$ is fixed throughout; the coupling between Fe2 and its next-nearest-Fe1 was fixed as $J_{22} = 0.4$; while the coupling between next-nearest-neighbor Fe1, which appears in the bulk case, was set to a realistic value of $J_{12} = 0.6$, see the text and the Supplementary materials for details. The dashed curves are the bulk $C_{\langle 11 \rangle}$ under these two conditions, as illustrated in the upper right insets. Solid curves are $C_{\langle 11 \rangle}$ of 1D chains of various length $L$ along the square edge (lower left inset of (a) with $L = 4$ (red), 6 (orange), 8 (pink), 10 (green) and 12 (blue) Fe1 sites and one Fe2 moment located in the middle interstitial site. The discontinuity in the curves corresponds to quantum phase transitions. The AFM correlation along the edge is systematically stronger than that in the bulk regardless of the $L$ chosen. (c) The $C_{\langle 11 \rangle}$ curves of a 1D open chain with two Fe2 moments along the square edge. The first Fe2 spin is fixed and the second Fe2 spin is placed at a distance of $x = 1$ (red), 2 (green), 3 (blue), and 4 (black) Fe1 sites from the first one (inset). The second Fe2 spin is unfixed. The corresponding case with only the fixed Fe2 spin is shown for comparison (purple dashed line). (d) Similar to (c) except that the second Fe2 spin is fixed to be parallel to the first Fe2 spin. These curves show a noticeable enhancement of AFM correlation between neighboring Fe1 sites when the Fe2 spins are parallel, i.e., ferromagnetically correlated, for the range of parameter of relevance.

If two Fe2 moments appeared in close proximity on the edge, our simulation showed that their ferromagnetic correlation would further enhance antiferromagnetic $C_{\langle 11 \rangle}$ (Fig. 4c and d). The dependence of $C_{\langle 11 \rangle}$ on the local Fe2 density could explain the inhomogeneous superfluid density on the edge which we observed (Fig. 2b). The spin system of FST composed of itinerant spins (Fe1) and magnetic impurities (Fe2) also bore resemblance to the Kondo lattice in heavy fermion superconductors where long range magnetic order might emerge [26] [48]. More interestingly, recent theoretical study based on the Hubbard model [49] suggested that edge ferromagnetism might indeed occur as a result of topological superconductivity in the bulk, which was possible because of the strong spin-orbit coupling inherent in FST superconductors [15] [16] [17] [21] [24]. It is therefore tantalizing to look for signs of long range ferromagnetic ordering on the edge which may couple to superconductivity in a potentially one-dimensional setting [50].

The "non-topological" mechanism we proposed above involves interesting interplay between spin physics and unconventional superconducting pairing interaction in FeSC [25]. Their connection with and distinction from the topological regime are worth further exploration. But regardless of the microscopic physical mechanism, our observation of robust superconducting edge channels in FST under strong magnetic perturbation bodes well for its application in superconducting spintronics where mitigating the negative effect of ferromagnetism on superconductivity is a challenge [29] [30].

In conclusion, we used sSQUID microscopy to find that the edge of "non-bulk" superconducting FST samples showed much stronger diamagnetism on the edge than in the bulk. While the $T_c$ and superfluid density of the bulk were both strongly suppressed by the magnetic perturbation of interstitial Fe in this system, the superfluid density of the edge was much higher and the $T_c$ of it was 14 K, the same as "bulk" superconducting FST samples in which the magnetic perturbation was absent. The edge also dominantly carried supercurrent biased into the sample at the temperature below $T_c$, suggesting a larger critical current than its bulk. Our Heisenberg modeling showed that antiferromagnetic spin correlation, which was believed to be vital to bulk superconductivity that could be strongly suppressed by Fe2 spins, was robust against such perturbation on the edge. Our result has implications for both understanding of the unconventional superconductivity of FeSC and future applications of superconductors in information technology.

**Acknowledgments**

Yihua Wang would like to acknowledge partial support by the Ministry of Science and Technology of China (2016YFA0301002 and 2017YFA0303000), the National Natural Science Foundation of China (11827805), and Shanghai Municipal Science and Technology Major Project Grant No.2019SHZDZX01. Da Jiang would like to acknowledge partial support by the "Strategic Priority Research Program (B)" of the Chinese Academy of Sciences (XDB04040300), the National Natural Science Foundation of China (11274333), and Hundred Talents Program of the Chinese Academy of Sciences. Shaoyu Yin would like to acknowledge support by National Natural Science Foundation of China (11704072). Work at Stanford was supported by an NSF IMR-MIP (DMR-0957616), and as part of the National Nanotechnology Coordinated Infrastructure under award ECCS-1542152. All the authors are grateful for the experimental assistance by Yuanbo Zhang, Shiyan Li, Donglai Feng and Changlin Zheng.